\documentstyle[11pt,paspconf,epsf,psfig,twoside]{article}
\begin{document}

\title{The 1996 dwarf nova outburst of the intermediate polar GK Persei}
\author{L.\,Morales-Rueda$^{1}$ M.\,D.\,Still$^{2}$ and
  P.\,Roche$^{1}$}
\affil{$^{1}$Astronomy Centre, University of Sussex, Falmer, Brighton, U.K.\\
  $^{2}$Physics and Astronomy, University of St Andrews, St Andrews, U.K.}

\begin{abstract}
  GK~Persei is an intermediate polar that exhibits a dwarf nova
  outburst approximately every 1000 days.  During previous outbursts
  the system has displayed kilo-second quasi-periodic oscillations
  (QPOs) in X-ray emission.  These oscillations are unique in that
  they are an order of magnitude longer than QPOs usually observed in
  disc-fed cataclysmic variables.  We have detected the optical
  counterpart of these QPOs in spectrophotometry collected during the
  1996 outburst.  The presence of these QPOs in Doppler-broadened
  emission lines originating from the highly-supersonic accretion flow
  enables us to place useful constraints on the location of the QPO
  source and its formation mechanism.
\end{abstract}


\section{Introduction}
GK~Per is an intermediate polar -- a cataclysmic variable, where the
magnetic field of the white dwarf is strong enough to truncate the
inner accretion disc if such a disc is at all present. Material
accreting through the disc is threaded onto the field lines, forced
out of the orbital plane along the accretion curtains and deposited
directly onto the magnetic poles of the white dwarf.  This system is
highly variable over several timescales -- the orbital period of the
binary -- 2-d (Crampton, Cowley \&\ Fisher 1986), and the spin period
of the white dwarf and accreting curtains -- 351-s (Ishida et\,al.
1992). In addition, the outburst cycle is caused by thermal
instabilities within the accretion disc, occurring over a $\sim$1000-d
timescale, and variability intermediate between orbit and spin is
provided by the Keplerian frequencies of gas in the disc.

We observed GK~Per in February 1996 (Morales-Rueda, Still \&\ Roche
1996) during one of its outburst states.  A frequency search across
the strong emission lines of H$\beta$ and He{\sc ii} revealed signal
at the spin frequency and 5-6~cycles/d window aliases (Fig.~1).
However, the strongest signal occurs at $\sim$5000-s -- the optical
counterpart of the X-ray QPOs observed in a previous outburst by
Watson, King \&\ Osborne (1985).

Watson et\,al. (1985) suggested that the X-ray QPOs were the
consequence of modulated mass transfer through the
magnetically-threaded accretion curtain -- where 5000-s is the beat
between the spin frequency and the Keplerian frequency of dense blobs
of gas orbiting close to the inner rim of the disc.

Hellier \&\ Livio (1994) proposed an alternative mechanism, where the
accretion stream from the companion star does not dissipate in a shock
at the outer disc edge, as is usually assumed.  Instead it overflows
the upper and lower disc surfaces and deposits blobs at its point of
closest approach to the white dwarf where it finally crashes into the
disc (Lubow \&\ Shu 1975; Armitage \&\ Livio 1996).  Blobs have higher
vertical structure and larger column densities than the surrounding
disc, and will have a Keplerian frequency of $\sim$5000-s.  The X-rays
emitted by shock-heated material from where the curtains impact the
white dwarf poles illuminate the disc and the blobs.  Consequently,
obscuration of the central X-ray source by the vertically-extended
blobs provides the X-ray QPOs, whilst, if we adopt this scenario,
X-ray reprocessing off the blobs provides the optical counterparts to
these oscillations. This model also explains the X-ray hardness ratio
measured from the data collected by Watson et\,al.  (1985) which is
consistent with the photo-electric absorption of soft X-rays by cool
intervening gas.

However, Fig.~1 shows that the optical QPOs are not symmetrically
distributed across the emission profiles. The blue wings of the lines
provide more power than the red wings at the QPO frequency. This
asymmetry is also obvious in the line profiles when binned into the
QPO period (see Fig.~2 of Morales et\,al. 1996), and is independent of
the orbital phase as seen in the trails presented in Morales-Rueda,
Still \&\ Roche (1998). We find this observation impossible to explain
in terms of the accretion stream overflow model.

\begin{figure}
\psfig{file=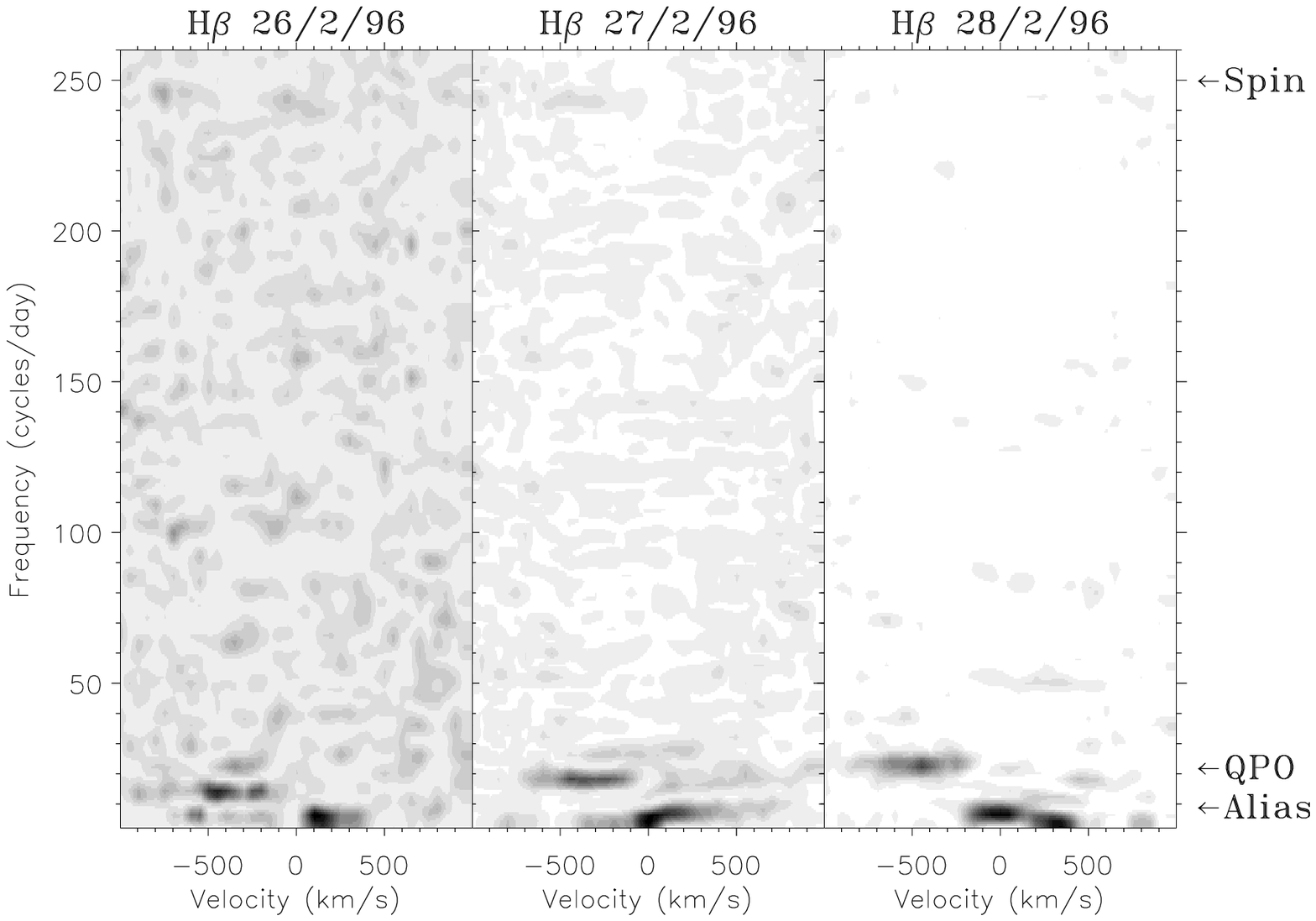,angle=90,height=8cm,width=14cm}
\psfig{file=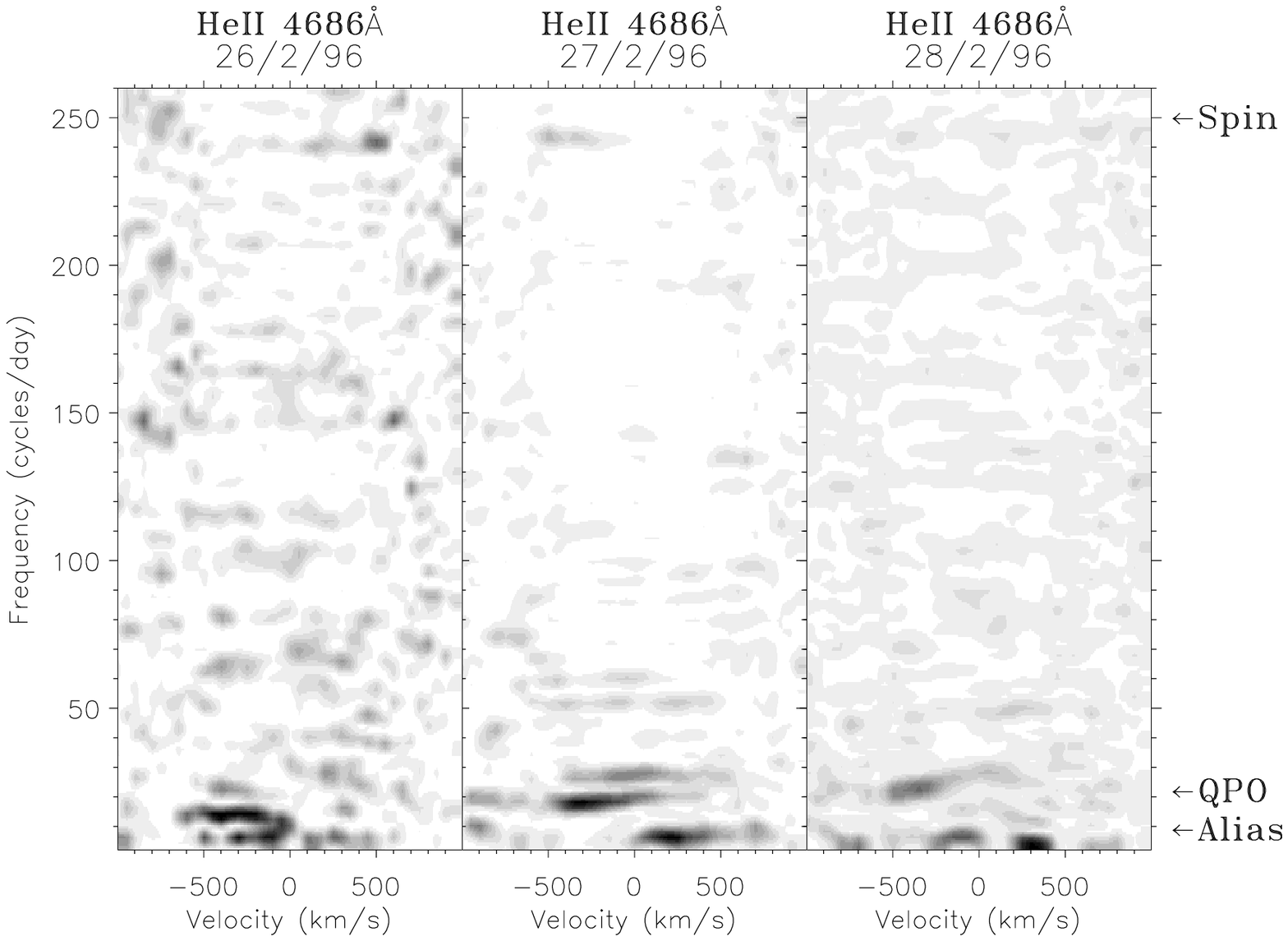,angle=90,height=8cm,width=14cm}
  \caption{Power spectra as a function of
    line velocity in the Doppler-broadened emission profiles of
    H$\beta$ and He{\sc ii}\,$\lambda$4686\AA, where negative and
    positive velocities represent the blue and red wings of the line
    respectively. We see power at the spin frequency of the white
    dwarf (246~c/d), and low-frequency observational aliases
    (5-6~c/d). However the power spectra are dominated by QPO signal
    at $\sim$\,18~c/d, where we see a blue--red asymmetry in the peak.}
\end{figure}

\section{A beat model incorporating shear line broadening}
Consider a dense blob of gas close to the inner edge of the accretion
disc and orbiting at a Keplerian frequency which will be close to the
co-rotation radius of the magnetic curtain.  On the order of every few
thousand seconds the base of the magnetic curtains will sweep over the
blob, increasing the density of material passing through the threading
regions and providing a cool absorbing body for the X-ray QPOs. We
expect to see modulation in the line intensity on the beat period
between the 351-s white dwarf spin and the orbital frequency of the
inner accretion disc.

Now consider the schematic of GK~Per provided in Fig.~2.  For this
figure, we take the orbital inclination to be consistent with
46$^{\circ} < i <$ 72$^{\circ}$ (Reinsch 1994). We have assumed that
the magnetic axis of the white dwarf is misaligned with the rotational
axis of the system by 45$^{\circ}$, although its true inclination is
unknown.

In Morales-Rueda et\,al (1998) we see that, similar to the X-ray QPOs,
this optical counterpart in the emission lines is the result of
absorption. There are at least two possible mechanisms which can
modulate the line emission by absorption over the beat cycle.  First
consider the observed absorption to be of line flux from the accretion
curtain. At the point in the spin cycle we label $\phi_{\rm spin} =
0$, the velocity gradient across the emitting surface behind the white
dwarf attains a maximum and the region is blue-shifted in the rest
frame.  The surface in front of the accretor is redshifted and has the
same velocity gradient, but is obscured from Earth by the inner edge
of the accretion disc.  Half a spin cycle later, at $\phi_{\rm spin} =
0.5$, both surfaces are equally visible but their velocity gradients
are at a minimum.  Horne \&\ Marsh (1986) have shown that in the
saturated case the strength of the absorption of a line element
increases $\sim$ linearly with the velocity gradient across the line
source.  The effects of shear broadening require maximum absorption to
occur when the upper accretion curtain and the blob are aligned behind
the white dwarf -- consistent with observing the majority of QPO power
in the blue wings of the emission lines.  A QPO period of 5000-s
suggests that blobs are orbiting the white dwarf with a period of
380-s or 320-s.  Power at 380-s has been detected previously in
optical photometry (Patterson 1981).

A second mechanism could be that the absorption is of continuum light
from the accretion disc behind the white dwarf.  Continuum absorption
by the upper accretion curtain will occur when it is behind the white
dwarf i.e.  once per spin cycle. At other spin phases this curtain has
no back-illuminating source.  The lower curtain has no
back-illuminating source at any time during the spin cycle. This
provides a natural blue asymmetry in the emission line signal.

\begin{figure}
\psfig{file=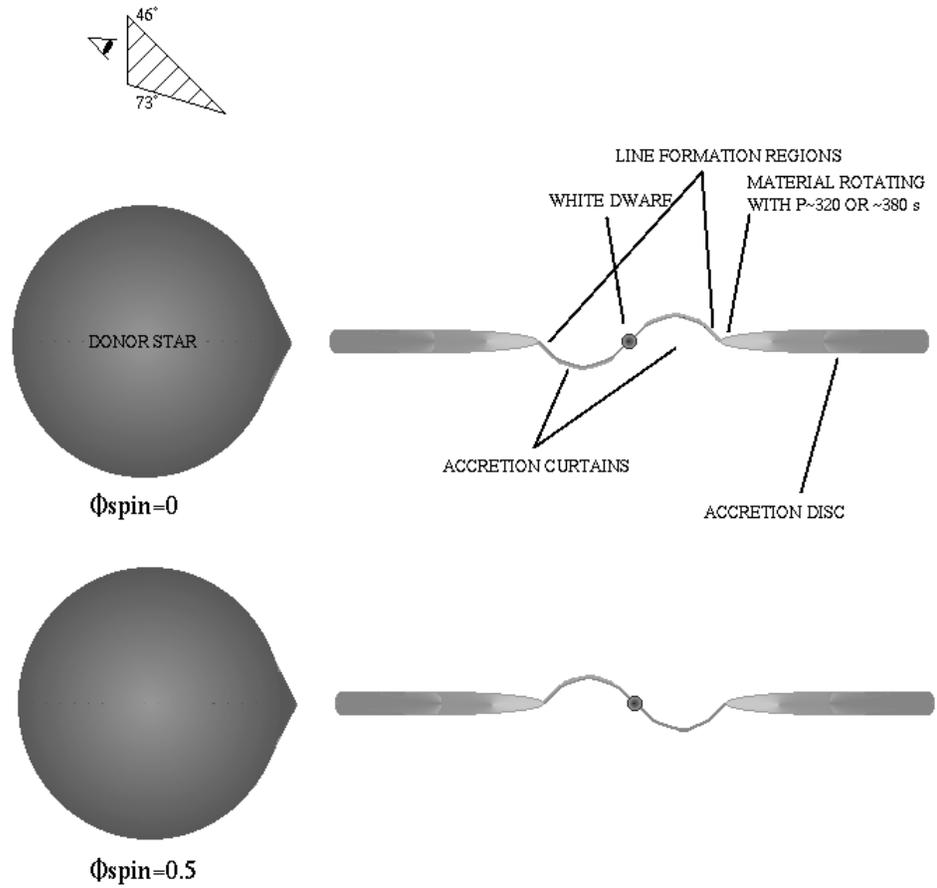,width=13cm}
\caption{Schematics of the accretion flow in the intermediate polar
  GK~Per. The position of the observer corresponds to orbital phase
  zero. The two diagrams are slices through a vertical plane across
  the binary line of centres, separated in time by half a spin cycle.}
\end{figure}

The beat mechanism explains the long-timescale QPOs from GK~Per using
pre-existing models of QPO generation. The driving mechanism has a
time-scale of a few hundred seconds in agreement with the QPOs
observed in other dwarf novae. In the above discussion we have
considered the QPO in terms of a blob orbiting in the inner accretion
disc but the alternative models of radially-oscillating acoustic waves
in the disc work equally well (Okuda et.\,al 1992).

\section{Conclusions}
The detection of kilo-sec QPOs across the optical emission line
profiles of GK~Per have provided the opportunity to test the mechanism
behind the unique long-timescale QPOs in this object, which are an
order of magnitude longer than QPOs normally observed in
disc-accreting cataclysmic variables.  We have rejected the model of
Hellier \&\ Livio (1994) considering the direct effects of blobs
orbiting at the Keplerian frequency of the annulus associated with a
disc-overflow impact site.  Our favoured model considers the long QPO
period to be the consequence of beating between ``usual'' timescale
QPOs at $\sim$ 300\,s with the magnetic accretion curtain spinning
with the white dwarf.  As a result of this, we do not require a new
model to explain these long timescales: the curiously-long
oscillations are merely a consequence of the magnetic nature of the
binary.

\acknowledgments We thank Tom Marsh for making his reduction software
available, and the Nuffield Foundation for a grant to PR in order to
facilitate this collaborative work. The reduction and analysis of the
data were carried out on the Sussex node of the STARLINK network. Ths
Isaac Newton Telescope is operated on the island of La Palma by the
Isaac Newton Group in the Spanish Observatorio del Roque de los
Muchachos of the Instituto de Astrof{\,i}sica de Canarias.

\end{document}